%% file: main.tex
\documentclass[sigconf,nonacm]{acmart}

\usepackage{algorithm}
\usepackage{algpseudocode}

\usepackage{listings}

\AtBeginDocument{%
  }

\begin{document}

\title{Serverless platform driven CPU loadbalancing}
%% \date{Oct 2025}

\author{Abdul Rehman}
\affiliation{%
  \institution{Indiana University Bloomington}
  \city{Bloomington}
  \state{Indiana}
  \country{USA}
}
\email{abrehman@iu.edu}
\orcid{0009-0005-1568-8068}

\renewcommand{\shortauthors}{Abdul et al.}

\begin{abstract}
\input{abstract}
\end{abstract}


%\ccsdesc[500]{Do Not Use This Code~Generate the Correct Terms for Your Paper}
%\ccsdesc[300]{Do Not Use This Code~Generate the Correct Terms for Your Paper}
%\ccsdesc{Do Not Use This Code~Generate the Correct Terms for Your Paper}
%\ccsdesc[100]{Do Not Use This Code~Generate the Correct Terms for Your Paper}



\maketitle

\input{introduction}
\input{background}

\input{motivation}

\input{custom_scheduler}

\input{loadbalancing_policy}

\input{implementation}

\input{evaluation}

\input{related_work}

\input{conclusion}

% \begin{acks}
% \end{acks}

\bibliographystyle{ACM-Reference-Format}
%\bibliography{sample-base}
\bibliography{refs.bib}


\end{document}

%% file: abstract.tex
Serverless platforms maintain a global view of function invocations and resource utilization, yet existing systems largely restrict CPU scheduling decisions to the operating system scheduler. This paper presents a serverless platform-driven CPU load balancing framework that enables the control plane to directly influence CPU scheduling through a custom Linux scheduler built on SchedExt (SCX). The proposed scheduler introduces configurable scheduling domains and a shared interface that allows the control plane to assign functions to domains based on their historical inter-arrival times. Within each domain, a single-queue load-balancing strategy combined with a virtual-time prioritization policy improves task placement while reducing interference from busy-polling tasks. Results show that an eight-domain configuration achieves the best trade-off, reducing system energy consumption by approximately 15\% while increasing invocation cost by only 5\%. Under heavily loaded workloads, the proposed scheduler also reduces request latency by up to 50\% compared to the default Linux Completely Fair Scheduler (CFS). These results demonstrate that exposing CPU scheduling decisions to the serverless control plane can improve both energy efficiency and workload performance while preserving scheduling flexibility.

%% file: introduction.tex
\section{Introduction}

Function-as-a-Service (FaaS) has become one of the dominant cloud computing models for deploying event-driven applications. By abstracting infrastructure management from developers, serverless platforms automatically provision resources, scale applications, and schedule function executions in response to workload demand \cite{aws}\cite{gcf}\cite{azure}. This programming model greatly simplifies application development while allowing cloud providers to optimize resource utilization across large-scale clusters. Consequently, the serverless control plane has evolved into a sophisticated resource manager responsible for scheduling functions, provisioning execution environments, and maintaining service-level objectives (SLOs) for latency and throughput.

CPU scheduling plays a central role in achieving these objectives. Modern serverless workloads consist of heterogeneous functions with diverse execution characteristics, ranging from short-lived I/O-bound requests to long-running CPU-intensive computations. Existing serverless platforms rely almost exclusively on the operating system scheduler, typically Linux's Completely Fair Scheduler (CFS), to allocate CPU time among executing functions. While the serverless control plane decides which worker machine executes a function, it has little influence over how CPU resources are shared once execution begins. This separation leaves potentially valuable workload information unavailable to the operating system scheduler.

The serverless control plane maintains rich historical information about function behavior, including invocation frequency, inter-arrival time, execution duration, and resource requirements. Existing research primarily exploits this information for cluster-level decisions such as worker placement, autoscaling, and container management \cite{autopilot}\cite{k8sautoscale}. In contrast, CPU scheduling remains a local operating system concern, driven solely by instantaneous runnable tasks. As a result, the scheduler cannot exploit application-level knowledge to improve cache locality, reduce interference between unrelated functions, or prioritize latency-sensitive workloads.

This paper argues that CPU scheduling should become a first-class responsibility of the serverless control plane. Rather than replacing the operating system scheduler, we expose a lightweight interface through which the control plane can influence CPU load balancing while allowing the kernel scheduler to retain responsibility for efficient task execution. Recent advances in Linux, specifically the SchedExt (SCX) framework and eBPF, make this practical by enabling custom scheduling policies without modifying the kernel.

We present a serverless platform-driven CPU load balancing framework that combines global workload knowledge from the serverless control plane with fine-grained scheduling decisions inside the Linux kernel. Our design partitions processors into configurable scheduling domains and allows the control plane to assign functions to these domains based on historical workload characteristics. A custom SCX scheduler then performs work-conserving single-queue load balancing within each domain using a virtual-time prioritization policy that favors short-lived, frequently sleeping tasks while de-prioritizing busy-polling tasks. This design improves processor locality, reduces interference among unrelated functions, and enables workload-aware CPU scheduling without sacrificing scheduling flexibility.

We implement the proposed framework by extending the Ilúvatar serverless platform with a custom eBPF scheduler built using SchedExt. Evaluation on a 48-core Intel Xeon server using traces sampled from the Microsoft Azure Functions dataset\cite{azuretrace} demonstrates that the proposed approach can substantially improve system efficiency. Compared to Linux CFS, our scheduler reduces energy consumption by approximately 15\% while increasing invocation cost by only 5\% in the best-performing configuration. Under heavily utilized workloads, it also reduces request latency by up to 50\%, illustrating the benefits of coordinating serverless resource management with operating system CPU scheduling.

The main contributions of this paper are:

\begin{itemize}
    \item  We identify the lack of coordination between serverless control planes and operating system CPU schedulers as a missed optimization opportunity for Function-as-a-Service platforms.
    \item We design an eBPF-based CPU scheduler that exposes a lightweight interface allowing the serverless control plane to steer CPU load balancing through configurable scheduling domains.
    \item We propose an inter-arrival-time-based scheduling-domain assignment policy that exploits historical function behavior to improve processor locality and energy efficiency.
    \item We implement the complete system in the Ilúvatar serverless platform using Linux SchedExt and evaluate it on realistic Azure serverless traces, demonstrating significant improvements in energy efficiency and performance compared to Linux CFS.
\end{itemize}

The remainder of the paper is organized as follows. Section 2 reviews the background on serverless computing, Linux CPU scheduling, eBPF, and the SchedExt framework. Section 3 motivates the need for serverless platform-driven CPU load balancing and discusses the limitations of existing scheduling interfaces. Section 4 presents the design of our custom CPU scheduler, while Section 5 describes the control-plane policy used to assign functions to scheduling domains. Section 6 details the implementation of our prototype in the Ilúvatar serverless platform using Linux SchedExt. Section 7 evaluates the proposed approach using microbenchmarks and Azure serverless workload traces. Section 8 discusses related work, and Section 9 concludes the paper and outlines directions for future research.

%% file: background.tex
\section{Background}

\subsection{Function as a Service (FaaS)} 
FaaS allows user to register small snippets of function code that get executed in response to an event (such as an HTTP request, message queue event, etc.) \cite{azure}\cite{aws}. These functions are stateless: a new execution environment is created for every invocation (and can be destroyed after the function returns). The function code contains all the necessary dependencies (imported libraries and packages), which significantly increases the initialization time before the event-handling code can execute. Functions are executed inside virtual execution environments such as hardware virtual machines, OS containers like Docker, or language based runtimes (javascript, WASM). Initialized function sandboxes can be retained in memory, and this keep-alive provides faster “warm-starts”. Since functions are arbitrary usercode, they are heterogeneous in their execution characteristics and resource requirements.

\subsection{FaaS Control Plane} 
All aspects of function execution are orchestrated by a serverless control plane, which are implemented by frameworks like OpenWhisk, Iluvatar, AWS Lambda, Google Cloud Functions \cite{openwhisk}\cite{iluvatar}\cite{aws}\cite{gcf}. For using a FaaS service, the user interacts with the control plane for registering and invoking functions, tracking their status, etc. The control plane manages the resources of a cluster of servers, and schedules functions on to them based on its load-balancing policies. Control plane has a global view of the cluster and can make load balancing decisions based on the resource utilization of
the cluster. We use Iluvatar\cite{iluvatar} for our setup because it provides a worker centric design for each server in the cluster. It allows us to implement worker driven CPU loadbalancing policy. 

\subsection{Linux Scheduling and Control Group Subsystem}
Linux has multiple scheduler classes as outlined by POSIX standard \cite{mansched}. For tasks belonging to a serverless function, normal scheduling class is used. It provides SCHED\_OTHER, SCHED\_IDLE and SCHED\_BATCH policies. SCHED\_OTHER is the default policy used by tasks of different containers. It provides CPU bandwidth control for control group subsystem to limit the CPU time given to a group of tasks. It is exposed through sysfs settings cpu\_quota and cpu\_period for each control group with cpu controller attached to it. Setting quota equal to period would allow the group of tasks to use one CPU worth of time each period. 

Linux also has realtime scheduling class which provides SCHED\_FIFO, SCHED\_RR and SCHED\_DEADLINE policies. Real time scheduling class allows static priority based scheduling wherein higher priority tasks can preempt lower priority tasks. Naturally, they can also preempt normal tasks. Starting from kernel 6.12 Linux has introduced a new scheduling class SCHED\_EXT below SCHED\_OTHER. It allows eBPF programs to implement arbitrary scheduling logic. 

\subsection{Extended Berkeley Packet Filter (eBPF)}

eBPF is a Linux kernel subsystem that allows safe execution of custom programs inside the Linux kernel at runtime \cite{ebpf}. eBPF programs are compiled into a standardized bytecode and loaded into the kernel after verification for execution. They are event driven, triggered by various hooks in the kernel (e.g., system calls, network packets, tracing points). Originally designed for network packet filtering, eBPF has evolved into a general-purpose framework used for observability, security, and performance monitoring. 

SchedExt framework uses struct\_ops eBPF programs to plug custom scheduling logic into the kernel \cite{schedext}. eBPF programs have different types depending on the attachment points. There are 30 program types listed in uapi/linux/bpf.h. XDP is one of them. Such programs are attached to a networking interface and help inspect ingress and egress packets. Struct\_ops is another program type that allows eBPF programs to register callbacks for predefined structures. 

\subsection{SchedExt (Scx)}
Scx is an eBPF based framework introduced in Linux kernel 6.12 \cite{schedext}. It introduced a new scheduling class below SCHED\_OTHER that allows eBPF programs to implement custom scheduling logic. User builds a struct\_ops type eBPF program that is attached to SchedExt core scheduling class. Operations provided by the struct\_ops define logic for various callbacks triggered by the scheduling class. 

SchedExt core consists of per CPU local DSQs and a global DSQ. Both of these queues are FIFO. Whenever CPU is ready to take a task at SchedExt priority level, it looks up local DSQ and global DSQ. If they are empty, dispatch callback of the eBPF scheduler is called. It is like a request to the eBPF scheduler to push a task to local DSQ. eBPF scheduler can also maintain it’s own custom DSQs which can function as a priority or a FIFO queue. Priority queues are ordered by the vtime passed during the enqueue operation.

Using custom DSQs and set of callbacks, arbitrary scheduling logic can be implemented. Production ready schedulers have been written using this framework. They run as a regular Linux process.  We employ SchedExt to build a custom scheduler driven by the FaaS control plane.

%% file: motivation.tex
\section{Motivation}

Serverless control plane has global view of the cluster and incoming function requests. It accumulates historic data about function types and their statistics. It can use this data for intelligent decision making. There is plethora of work on using this data for cluster loadbalancing but using this data for cpu loadbalancing is an unexplored area.

Serverless control plane can use historic statistics about function like inter arrival time to group together tasks of high frequency functions on few cores. This would promote core reuse by majority of the tasks across the workload. It has already been shown by Julia et al \cite{nest} that reusing cores improve the energy efficiency of the system. They have implemented this idea at the scheduler level by remembering last used core of the task. Instead, in the context of serverless, the control plane can implement this idea in a more proactive manner by looking at the historic statistics of the functions.

To drive CPU loadbalancing based on inter arrival time of the function, the serverless control plane needs a mechanisim to be able to affect CPU loadbalancing. Existing CFS interface does not provide any such mechanisim. It provides an interface to limit CPU time through CFS bandwidth control and define CPU affinity of the tasks (pinning tasks to cores). This motivates the requirement for a custom scheduler with an interface for the control plane to affect the CPU loadbalancing.

The custom scheduler with this new requirement has to maintain existing invariants of the CFS. The most important of them being the mechanism to limit CPU usage among functions and the notion of scheduling domains for loadbalancing. Linux defines scheduling domains over logical CPU, caches and Numa nodes. 

We allow the control plane to define scheduling domains over arbitrary set of cores that are mutually exclusive. This simplifies the scheduler design and moves the responsibility of limiting CPU usage among functions from scheduler to the control plane. The control plane can limit the CPU usage among functions by limiting the number of functions assigned to a scheduling domain.

%% file: custom_scheduler.tex
\section{Custom Scheduler}
\label{sec:customscheduler}

\begin{figure}[htbp]
     \centering
     \includegraphics[trim=0 50 170 0,clip,width=0.9\linewidth]{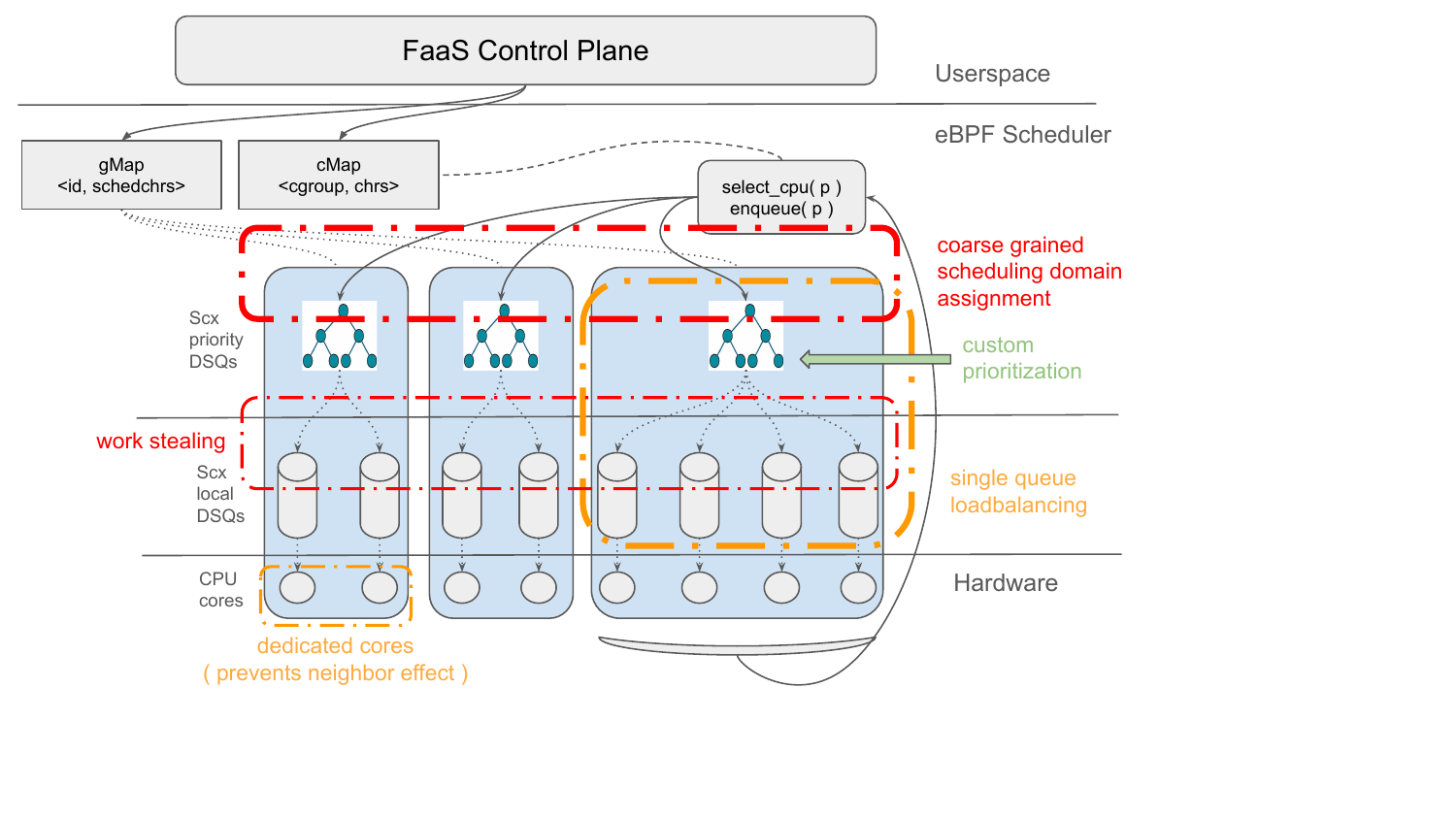}
     \caption{CPU Scheduler Design}
     \label{fig:sched_design}
 \end{figure}

The custom scheduler provides an interface for the serverless control plane to define loadbalancing among the scheduling domains. It creates a shared eBPF map which the serverless control plane populates with control group name of the function and it's assigned scheduling domain id. Scheduler consumes this shared map to enqueue tasks into priority queue of the respective scheduling domain. Whenever a task is forked, expires timeslice or is woken up, the scheduler get's it's scheduling domain id from the map and enqueue's it into the priority queue prioritized according to Algorithm-\ref{code:vtime_priority}. It implements single queue loadbalancing within a scheduling domain which is theoretical optimum. Next available CPU picks the task from the head of the queue, implementing work conservation within a domain - Algorithm-\ref{code:custom_scheduler}. To rebalance the scheduling domains, the serverless control plane updates the domain assignment in the shared map. Changes are reflected at each scheduling cycle for the task. Figure-\ref{fig:sched_design} outlines the design of the custom scheduler.

\begin{algorithm}
\caption{Vtime Prioritization}
\label{code:vtime_priority}
\begin{algorithmic}[1]
\State \textbf{Given:} $n$ to reset vtime to current time 
\State \textbf{Call when:} a task is being enqueued to a dsq
\Procedure{EnqueueTask}{$dsq,\ p$}
    \If{$p.enqueue\_count\ \%\ n==0$}
        \State $p.vtime = current\_time$
    \Else
        \State $p.vtime\ +=\ p.last\_cpu\_time/p.sleep\_frequency$
    \EndIf
    \State $p.enqueue\_count\  += 1$
    \State $p.slice  = default\_timeslice $
   \State $dsq.enqueue(p)$
\EndProcedure
\end{algorithmic}
\end{algorithm}

\subsection{Vtime Prioritization}

Single queue loadbalancing uses a priority queue prioritized based on vtime. Algorithm-\ref{code:vtime_priority} updates the vtime in a way to prioritize tasks which consume less cpu time and sleep often while maintaining fairness using current time every $n^{th}$ enqueue. This de-prioritizes busy polling tasks which consume whole timeslice and gives preference to shorter worker tasks.

\begin{algorithm}
\caption{eBPF Custom Scheduler}
\label{code:custom_scheduler}
\begin{algorithmic}[1]
\State \textbf{Call when:} a task is forked, woken up or expires timeslice 
\Procedure{EnqueueTask}{$p$}
    \If{$shared\_map[p\rightarrow cgroup]$}
        \State $preferred\_domain \leftarrow shared\_map[p.cgroup]$
        \State $preferred\_domain.dsq.enqueue(p)$
    \Else
        \State $global\_dsq.enqueue(p)$
    \EndIf
\EndProcedure
\
\State \textbf{Call when:} a cpu is empty 
\Procedure{PickNextTask}{$cpu$}
    \State $domain\_dsq \leftarrow cpu\_to\_domain\_dsq(cpu)$
    \State $task \leftarrow pick\_from\_head(domain\_dsq)$
    \
    \If{$task \cap \varnothing$}
        \State $task \leftarrow pick\_from\_head(global\_dsq)$
    \EndIf
    \
    \If{$task \cap \varnothing$}
        \State $task \leftarrow worksteal\_from\_neighbors()$
    \EndIf
    \
    \State \textbf{return} $task$
\EndProcedure
\end{algorithmic}
\end{algorithm}

%% file: loadbalancing_policy.tex
\section{Serverless platform driven loadbalancing}
\label{sec:loadbalancing_policy}

The serverless control plane uses the customer scheduler interface to drive the CPU loadbalancing. It assigns the function a scheduling domain and populates the shared eBPF map with assignment. Assignment takes affect at each scheduling cycle.

Scheduling domain assignment is made through a heuristic driven by function inter arrival time (IAT) - Algorithm-\ref{code:loadbalancing}. Whenever a function request arrives, the historic inter arrival time of the function is hashed using a predefined bucket\_size into a bucket\_id. Bucket\_id is used to lookup cached domains. An available domain is selected from among the cached domains. If none is found, a domain is selected from the whole set and cached into the bucket for future use.

Scheduling domain assignment heuristic creates group of domains that are reused by high frequency functions. This increases the likehood of tasks finding warm cores. It has been shown by Julia et al \cite{nest} that reusing warm cores is energy efficient. This heuristic loadbalance the tasks based on IAT of the functions to create warm domains for optimizing energy use.

\begin{algorithm}
\caption{IAT consistent hashing loadbalancing}
\label{code:loadbalancing}
\begin{algorithmic}[1]
\State \textbf{Given:} set of domains $D$ and required cpu count $cpus$
\Procedure{PickAvailableDomain}{$D,\ cpus$}
    \For{each domain $d$ in $D$}
        \If{can\_serve\_and\_acquire($d,\ cpus$)}
            \State \textbf{return} $d$
        \EndIf
    \EndFor
    \State \textbf{return} $\varnothing$
\EndProcedure
\
\State \textbf{Given:} set of domains in the system $D_{sys}$ and bucket\_size
\State \textbf{Call when:} a function invocation $f_r$ arrives 
\Procedure{AssignDomainToRequest}{$f_r$}
    \State $bucket\_id \leftarrow \lfloor f_r.IAT/bucket\_size \rfloor$
    \State $D_{preferred} \leftarrow domains\_map[bucket\_id]$
    \
    \State $d \leftarrow PickAvailableDomain(D_{preferred},\ fr.cpus)$
    \If{$d \cap \varnothing$}
        \State $d \leftarrow PickAvailableDomain(D_{sys},\ fr.cpus)$
    \EndIf
    \
    \State $domain\_map[bucket\_id] \leftarrow D_{preferred} \cup \{d\}$
    \State \textbf{return} $d$
\EndProcedure
\
\State \textbf{Call when:} a function invocation $f_r$ is complete  
\Procedure{RequestIsComplete}{$f_r$}
    \State $d \leftarrow f_r.domain$
    \State $return\_cpus\_and\_release(d,\ f_r.cpus)$
\EndProcedure
\end{algorithmic}
\end{algorithm}

%% file: implementation.tex
\section{Implementation}

We have implemented the serverless platform driven loadbalancing using a research oriented serverless control plane \textit{Iluvatar}\cite{iluvatar} and Scx framework\cite{schedext}. \textit{Iluvatar} make it easy to customize resource assignment part of the control plane while Scx allows writing custom CPU schedulers as userspace processes.

\subsection{Custom CPU Scheduler} 

Custom scheduler outlined in section-\ref{sec:customscheduler} is implemented using Scx framework. Scx framework provides \textit{struct\_ops} eBPF interface to allow userspace process to write custom logic for task scheduling. It implements an scx core as a lower priority CPU scheduling core than CFS to hook this logic into Linux scheduling core. Scx provides local dsqs (fifo queues) for each CPU. The Scx core picks task from these queues on each core at each scheduler tick if there are no other high priority tasks. Tasks are enqueued to local dsqs from the eBPF scheduler (struct\_ops logic). Scx provides various callbacks for the life cycle of the task. Select\_cpu is called when a task is being woken up. Enqueue is called when a task expires timeslice or is migrating. Dispatch is called when a local dsq becomes empty. eBPF scheduler can also create custom dsqs which are either priority queues or fifo queues. It can then move tasks from custom dsqs to local dsqs. 

Custom scheduler monitors the creation of new cgroups and switches the scheduling policy of tasks which belong to docker cgroup from CFS to Scx. This de-prioritizes the serverless function tasks from rest of system tasks and allows arbitrary scheduling logic to be implemented. It is responsible to read the shared map and schedule tasks within the preferred domains. It creates custom dsq for each domain and a global custom dsq in bpf land. It enqueues tasks into custom dsqs from enqueue callback. Tasks which belong to a cgroup with preferred domain in the shared map are enqueued to respective queue. Rest are enqueued to global queue. When local dsq becomes empty it moves tasks from custom dsq to local dsq of the cpu from dispatch callback. If the domain and global dsqs are empty it steals tasks from neighbors, starting from immediate neighbors and expanding the search one by one to the entire system on each side. This implements single queue loadbalancing. It is theoratical optimum and practical in this context because serverless tasks consume more than $1ms$ of cpu time per timeslice while the loadbalancing overheads are small $\approx10us$. Each custom dsq acts as a priority queue, prioritized by the vtime prioritization scheme outlined in section-\ref{sec:customscheduler}. 

\subsection{Serverless Control Plane driven loadbalancing} 

Iluvatar allows configuring each worker node. Each worker node is configured to use customized loadbalancing with a predefined set of domains. During initialization the worker creates a shared eBPF map with domain id and set of cores belonging to the domain for the custom scheduler to consume. Then it spawns the custom scheduler. The scheduler consumes this map and creates custom dsqs for the domains. 

Each incoming request to the worker is enqueued in a fifo queue before they are assigned resources for execution. A monitor thread picks the next request from the queue and assigns CPU resources. Loadbalancing policy outlined in section-\ref{sec:loadbalancing_policy} is implemented at this point. Cgroup created by container for the request is assigned a scheduling domain and another shared eBPF map is populated with this assignment. The request is then passed on to the container for execution. After the request is completed, domain is released but the shared map is left intact as an optimization.

%% file: evaluation.tex
\section{Evaluation}
\label{sec:evaluation}

\subsection{Setup}

\textbf{Hardware:} We use two set of servers, one for load generation and the other for iluvatar control plane worker. Worker machine has 48-core Intel Xeon Platinum 8160 CPUs with 1 TB of memory and 1 TB of NVMe storage. 

\textbf{Software:} Worker machine is setup with custom kernel 6.11.0-rc1 patched using SchedExt patches for Ubuntu 20.04. The kernel is configured using default kernel config for Ubuntu, having 250Hz scheduler tick and 3ms CFS timeslice. Custom scheduler also uses 3ms timeslice as default.

Each function runs in a docker container. Before the start of open loop workload generation, we create sufficient containers (prewarm stage) so that each invocation during the trace execution would be a warm invocation. It removes the cold start overheads and ensure a fair comparison.

\begin{figure}[htbp] % Use figure* for full page width (two-column documents)
    \centering
    \includegraphics[width=0.4\textwidth]{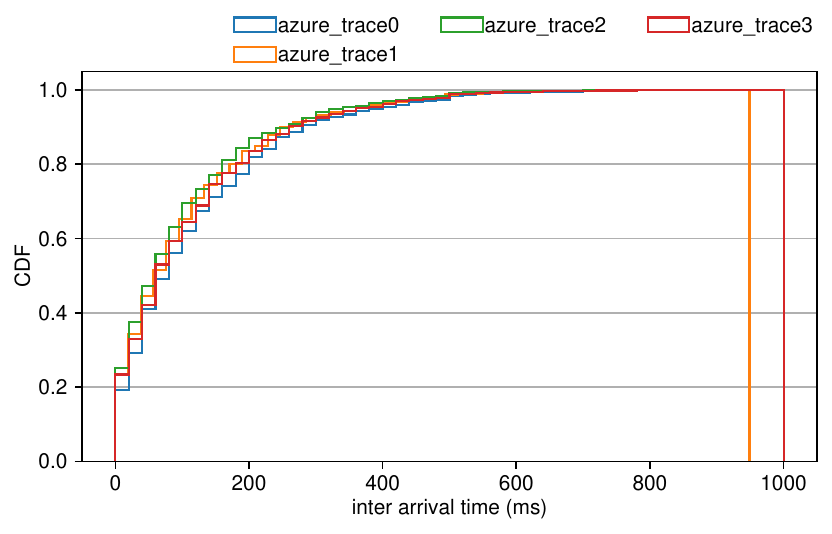}
    \caption{Cumulative density function of traces}
    \label{fig:eval:summary:traces_cdf}
\end{figure}

\textbf{Workload Generation:} We have sampled four different traces from Azure dataset with different number of functions generating a spectrum of system load. Table-\ref{tab:trace-characteristics} lists the number of functions in each trace along with average 1-min loadaverage of system running stock CFS. Figure-\ref{fig:eval:summary:traces_cdf} shows the cdf for the traces.

\begin{table}[h]
    \centering
    \caption{Trace Characteristics}
    \label{tab:trace-characteristics}
    \begin{tabular}{lcccccc}
        \hline
        name & $\lambda-$count & loadaverage(CFS) & Q-length & Saturation &  \\ \hline
        trace0 & 24               & 1.7 & 0             &  under-utilized                  &  \\
        trace1 & 28               & 65 & 0             &  medium-utilized                  &  \\
        trace2 & 29               & 170 & 0             &  medium-utilized                  &  \\
        trace3 & 30               & 722 & 1.9             &  over-utilized                  &  \\ \hline
    \end{tabular}
\end{table}

\subsection{Closed loop Microbenchmarks}

Microbenchmarking has been performed using closed-loop workload generator. It generates a single request and then waits for a response. Results are averaged over $200$ invocations. 

Four system setups have been benchmarked - CFS, 1domain, 8domains and 16domains. n-domains implies that system has been divided into $n$ scheduling domains - each having equal CPU count. Results are normalized with respect to CFS.

Custom scheduler performs equivalent to CFS for each function in microbenchmarks except for multithreaded functions (graph-pagerank and image-recognition) and single threaded long running cpu intensive function (video-processing). Multithreaded functions outperform CFS in 1domain and 8domain configurations while performing much worse in 16domain configuration. Video-processing performs worse than CFS for each configuration. 

\begin{figure}[htbp] % Use figure* for full page width (two-column documents)
    \centering
    \includegraphics[width=0.4\textwidth]{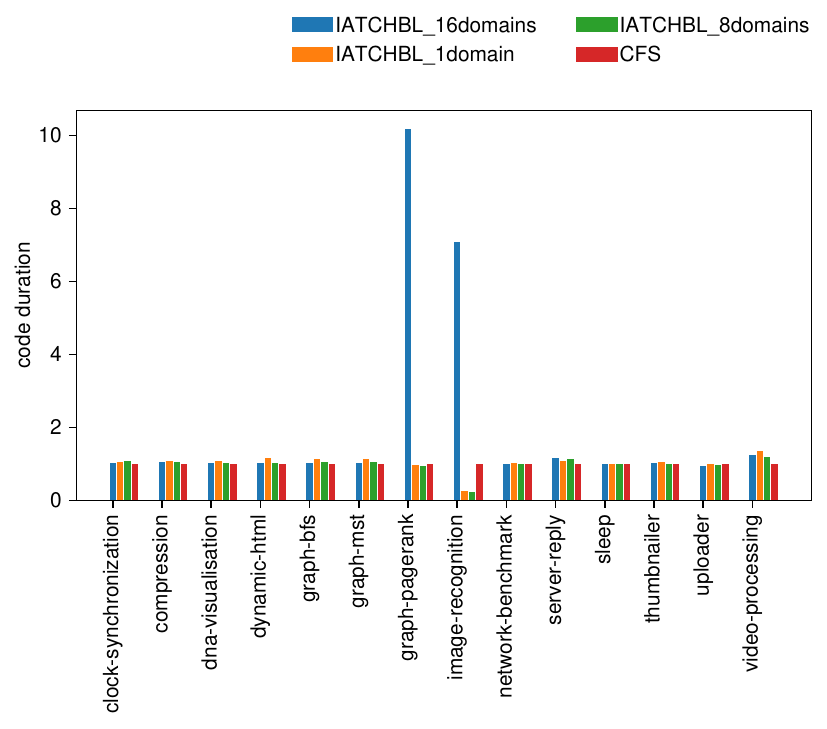}
    \caption{Closed loop microbenchmarks: code duration.}
    \label{fig:eval:microbenchmarks:code-dur}
\end{figure}

\begin{figure}[htbp] % Use figure* for full page width (two-column documents)
    \centering
    \includegraphics[width=0.4\textwidth]{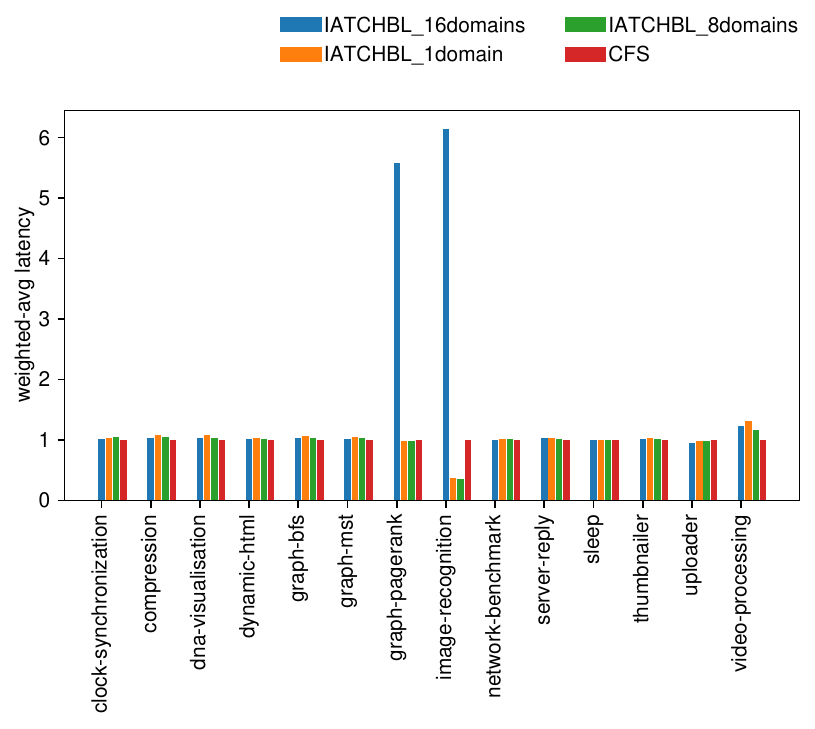}
    \caption{Closed loop microbenchmarks: weigted average latency.}
    \label{fig:eval:microbenchmarks:avg-lat}
\end{figure}

\begin{figure}[htbp] % Use figure* for full page width (two-column documents)
    \centering
    \includegraphics[width=0.4\textwidth]{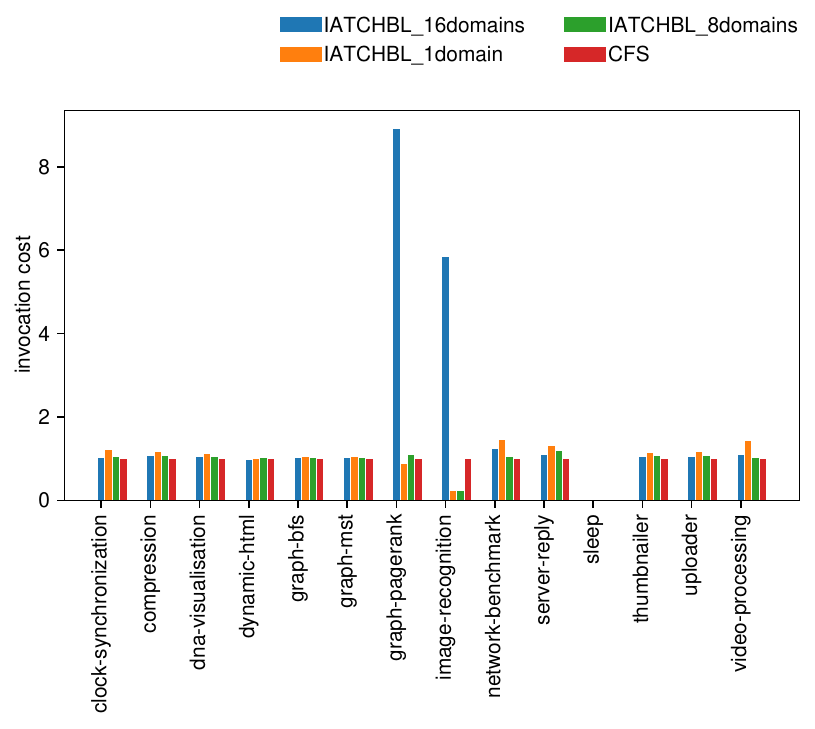}
    \caption{Closed loop microbenchmarks: invocation cost.}
    \label{fig:eval:microbenchmarks:invoc-cost}
\end{figure}

Figure-\ref{fig:eval:microbenchmarks:code-dur} shows comparison of code duration within the container. It excludes the control plane overheads. Figure-\ref{fig:eval:microbenchmarks:avg-lat} shows comparison of end to end latency and Figure-\ref{fig:eval:microbenchmarks:invoc-cost} compares invocation cost. Invocation cost is the product of latency and cpu utilization over the invocation - $invocation\_cost=latency*cpu\_utilization$.

Reason for eccentric results of multithreaded functions is inherent in custom scheduler design. It fulfills the CPU request by limiting the tasks to a scheduling domain. It limits the physical cpus available for multithreaded function. In contrast, CFS fulfills the request by limiting the cpu time for the cgroup in a given period. Our approach prevents neighbor effect caused by multithreaded serverless functions which is unavoidable by design in CFS. Consequently, function latency is drastically increased when a scheduling domain has too few cores - 3 cores per domain for 16 domains in a 48 core system. Whereas it outperforms for 1domain and 8domain configurations by virtue of single queue loadbalancing.

Single threaded long running cpu intensive function accumulates the overhead associated with the custom scheduler over the invocation. Video-processing adds a watermark to a video using ffmpeg in a single threaded process taking $0.74 secs$ with CFS. The latency increases by $\approx26.3\%$ with custom scheduler having a 3ms timeslice. In contrast, most functions in serverless workloads are multithreaded or ephemeral.

\subsection{Workload}

\begin{figure}[htbp] % Use figure* for full page width (two-column documents)
    \centering
    \includegraphics[width=0.4\textwidth]{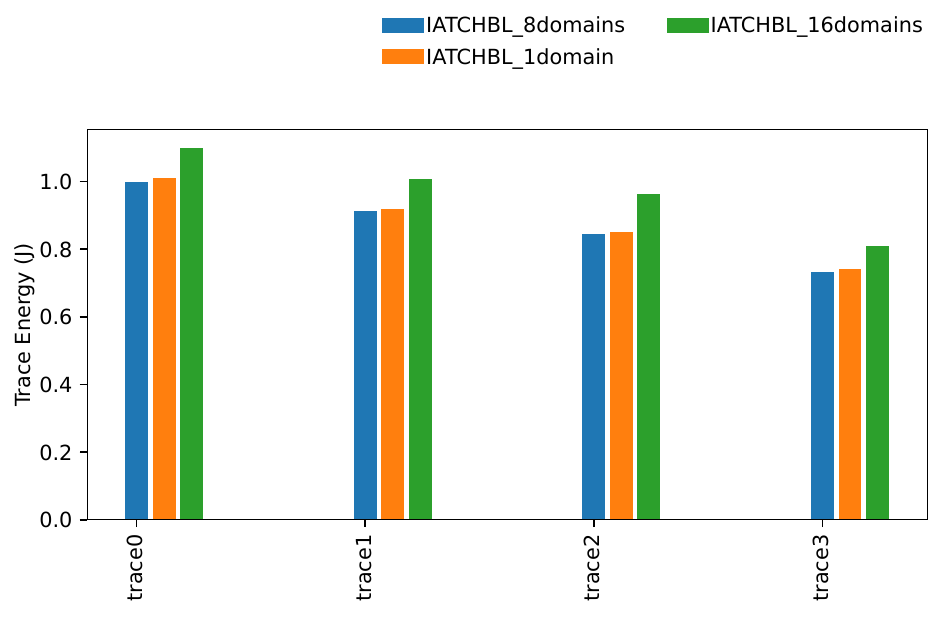}
    \caption{Workload: energy consumption}
    \label{fig:eval:summary:energy}
\end{figure}

Evaluation over traces sampled from Azure dataset indicate $\approx15\%$ energy saving as compared to CFS for mediumly utilized system - Figure-\ref{fig:eval:summary:energy}. Custom scheduler saves energy by prioritizing worker tasks over busy polling tasks. Worker tasks in serverless are mostly asleep and are only woken up on request for work. Busy polling tasks keep looping in a tight loop waiting for an event. In serverless, asynchronous frameworks implement state machine using busy polling to iterate over asynchronous functions. Such tasks consume whole timeslice without doing any useful work. 

Single queue in custom scheduler is a priority queue that prioritizes tasks that have high sleep frequency and consume less cpu time. CPU time distribution of CFS, 1domian and 16 domain showcase this difference. Listing-\ref{lst:distribution:cputime:cfs} show there are $91.3\%$ tasks that spend less then 1ms in CPU context. This number is increased with custom scheduler in 1domain configuration to $94.7\%$ - Listing-\ref{lst:distribution:cputime:1domain}. In 16domain configuration the energy saving benefit is gone and this number is also equivalent to CFS case - Listing-\ref{lst:distribution:cputime:16domain}. 16domain configuration becomes equivalent to CFS because tasks are divided among many queues just like CFS which has per core run queues. This prevents the single queue prioritization to reap benefits. 

\begin{lstlisting}[basicstyle=\tiny\ttfamily, caption={CPU time distribution: trace1,CFS}, label={lst:distribution:cputime:cfs}]
[2, 4)              0.013|@@@                                                 |
[4, 8)              0.206|@@@@@@@@@@@@@@@@@@@@@@@@@@@@@@@@@@@@@@@@@@@@@@@@@@@@|
[8, 16)             0.182|@@@@@@@@@@@@@@@@@@@@@@@@@@@@@@@@@@@@@@@@@@@@@       |
[16, 32)            0.115|@@@@@@@@@@@@@@@@@@@@@@@@@@@@                        |
[32, 64)            0.072|@@@@@@@@@@@@@@@@@@                                  |
[64, 128)           0.144|@@@@@@@@@@@@@@@@@@@@@@@@@@@@@@@@@@@@                |
[128, 256)          0.034|@@@@@@@@                                            |
[256, 512)          0.025|@@@@@@                                              |
[512, 1K)           0.048|@@@@@@@@@@@                                         |
[1K, 2K)            0.074|@@@@@@@@@@@@@@@@@@                                  |
[2K, 4K)            0.061|@@@@@@@@@@@@@@@                                     |
[4K, 8K)            0.008|@                                                   |
[8K, 16K)           0.006|@                                                   |
[16K, 32K)          0.004|@                                                   |
[32K, 64K)          0.003|                                                    |
[64K, 128K)         0.002|                                                    |
[128K, 256K)        0.001|                                                    |
\end{lstlisting}
\begin{lstlisting}[basicstyle=\tiny\ttfamily, caption={CPU time distribution: trace1,1domain}, label={lst:distribution:cputime:1domain}]
[2, 4)              0.000|                                                    |
[4, 8)              0.058|@@@@@@@@@@                                          |
[8, 16)             0.294|@@@@@@@@@@@@@@@@@@@@@@@@@@@@@@@@@@@@@@@@@@@@@@@@@@@@|
[16, 32)            0.146|@@@@@@@@@@@@@@@@@@@@@@@@@                           |
[32, 64)            0.093|@@@@@@@@@@@@@@@@                                    |
[64, 128)           0.153|@@@@@@@@@@@@@@@@@@@@@@@@@@@                         |
[128, 256)          0.041|@@@@@@@                                             |
[256, 512)          0.027|@@@@                                                |
[512, 1K)           0.050|@@@@@@@@                                            |
[1K, 2K)            0.087|@@@@@@@@@@@@@@@                                     |
[2K, 4K)            0.025|@@@@                                                |
[4K, 8K)            0.007|@                                                   |
[8K, 16K)           0.006|@                                                   |
[16K, 32K)          0.005|                                                    |
[32K, 64K)          0.004|                                                    |
[64K, 128K)         0.003|                                                    |
[128K, 256K)        0.001|                                                    |
\end{lstlisting}
\begin{lstlisting}[basicstyle=\tiny\ttfamily, caption={CPU time distribution: trace1,16domains}, label={lst:distribution:cputime:16domain}]
[2, 4)              0.000|                                                    |
[4, 8)              0.056|@@@@@@@@@@                                          |
[8, 16)             0.272|@@@@@@@@@@@@@@@@@@@@@@@@@@@@@@@@@@@@@@@@@@@@@@@@@@@@|
[16, 32)            0.145|@@@@@@@@@@@@@@@@@@@@@@@@@@@                         |
[32, 64)            0.093|@@@@@@@@@@@@@@@@@                                   |
[64, 128)           0.156|@@@@@@@@@@@@@@@@@@@@@@@@@@@@@                       |
[128, 256)          0.044|@@@@@@@@                                            |
[256, 512)          0.028|@@@@@                                               |
[512, 1K)           0.046|@@@@@@@@                                            |
[1K, 2K)            0.073|@@@@@@@@@@@@@@                                      |
[2K, 4K)            0.067|@@@@@@@@@@@@                                        |
[4K, 8K)            0.006|@                                                   |
[8K, 16K)           0.005|                                                    |
[16K, 32K)          0.004|                                                    |
[32K, 64K)          0.003|                                                    |
[64K, 128K)         0.002|                                                    |
[128K, 256K)        0.001|                                                    |
\end{lstlisting}

There are $\approx10\%$ busy polling tasks in the system on average in our setup. To identify these tasks, we sample the instruction pointer of the task at each scheduler tick. Then, we use the running mean of the instruction pointer to calculate it's running variance. Distribution of this variance for trace1, 1domain configuration is shown in Listing-\ref{lst:distribution:ipvar:1domain}. There are $11\%$ tasks with variance less then $4$ - instruction pointer stays within two instructions across scheduler ticks. De-prioritizing these $10\%$ tasks has improved the energy efficiency of the system by $\approx15\%$. 

\begin{lstlisting}[basicstyle=\tiny\ttfamily, caption={IP variance: trace1,1domain}, label={lst:distribution:ipvar:1domain}]
[0]                 0.092|@@@@@@@@@@@@@@@@@@@@@@@@@@@@@@@@@@@@@@@@@@@@@@@@@@@@|
[1]                 0.019|@@@@@@@@@@                                          |
[2, 4)              0.008|@@@@                                                |
[4, 8)              0.007|@@@@                                                |
[8, 16)             0.009|@@@@                                                |
[16, 32)            0.010|@@@@@                                               |
[32, 64)            0.016|@@@@@@@@                                            |
[64, 128)           0.013|@@@@@@@                                             |
[128, 256)          0.011|@@@@@@                                              |
[256, 512)          0.012|@@@@@@                                              |
[512, 1K)           0.013|@@@@@@@                                             |
[1K, 2K)            0.015|@@@@@@@@                                            |
[2K, 4K)            0.019|@@@@@@@@@@                                          |
[4K, 8K)            0.025|@@@@@@@@@@@@@@                                      |
[8K, 16K)           0.038|@@@@@@@@@@@@@@@@@@@@@                               |
[16K, 32K)          0.071|@@@@@@@@@@@@@@@@@@@@@@@@@@@@@@@@@@@@@@@             |
[32K, 64K)          0.053|@@@@@@@@@@@@@@@@@@@@@@@@@@@@@@                      |
[64K, 128K)         0.008|@@@@                                                |
[128K, 256K)        0.007|@@@@                                                |
[256K, 512K)        0.007|@@@@                                                |
[512K, 1M)          0.008|@@@@                                                |
[1M, 2M)            0.008|@@@@                                                |
[2M, 4M)            0.008|@@@@                                                |
[4M, 8M)            0.008|@@@@                                                |
[8M, 16M)           0.009|@@@@@                                               |
[16M, 32M)          0.010|@@@@@                                               |
[32M, 64M)          0.009|@@@@                                                |
[64M, 128M)         0.009|@@@@                                                |
[128M, 256M)        0.010|@@@@@                                               |
[256M, 512M)        0.013|@@@@@@@                                             |
[512M, 1G)          0.009|@@@@@                                               |
[1G, 2G)            0.008|@@@@                                                |
[2G, 4G)            0.009|@@@@@                                               |
[4G, 8G)            0.009|@@@@@                                               |
[8G, 16G)           0.010|@@@@@                                               |
[16G, 32G)          0.010|@@@@@                                               |
[32G, 64G)          0.011|@@@@@@                                              |
[64G, 128G)         0.011|@@@@@@                                              |
[128G, 256G)        0.012|@@@@@@                                              |
[256G, 512G)        0.013|@@@@@@@                                             |
[512G, 1T)          0.015|@@@@@@@@                                            |
[1T, 2T)            0.014|@@@@@@@@                                            |
[2T, 4T)            0.014|@@@@@@@@                                            |
[4T, 8T)            0.016|@@@@@@@@@                                           |
[8T, 16T)           0.017|@@@@@@@@@                                           |
[16T, 32T)          0.015|@@@@@@@@                                            |
[32T, 64T)          0.016|@@@@@@@@@                                           |
[64T, 128T)         0.018|@@@@@@@@@@                                          |
[128T, 256T)        0.019|@@@@@@@@@@                                          |
[256T, 512T)        0.018|@@@@@@@@@@                                          |
[512T, 1P)          0.015|@@@@@@@@                                            |
[1P, 2P)            0.016|@@@@@@@@                                            |
[2P, 4P)            0.017|@@@@@@@@@                                           |
[4P, 8P)            0.018|@@@@@@@@@@                                          |
[8P, 16P)           0.019|@@@@@@@@@@                                          |
[16P, 32P)          0.024|@@@@@@@@@@@@@                                       |
[32P, 64P)          0.024|@@@@@@@@@@@@@                                       |
[64P, 128P)         0.024|@@@@@@@@@@@@@                                       |
[128P, 256P)        0.023|@@@@@@@@@@@@@                                       |
[256P, 512P)        0.010|@@@@@                                               |
[512P, 1E)          0.000|                                                    |
[1E, 2E)            0.000|                                                    |
[2E, 4E)            0.000|                                                    |
[4E, 8E)            0.000|                                                    |
\end{lstlisting}

\begin{figure}[htbp] % Use figure* for full page width (two-column documents)
    \centering
    \includegraphics[width=0.4\textwidth]{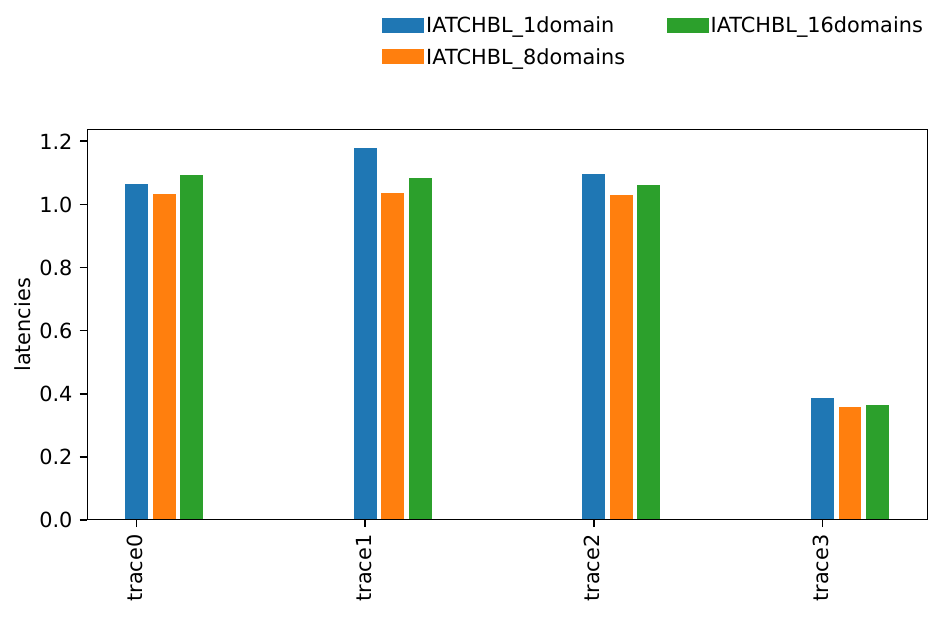}
    \caption{Workload: latencies}
    \label{fig:eval:summary:E2EMean}
\end{figure}

\begin{figure}[htbp] % Use figure* for full page width (two-column documents)
    \centering
    \includegraphics[width=0.4\textwidth]{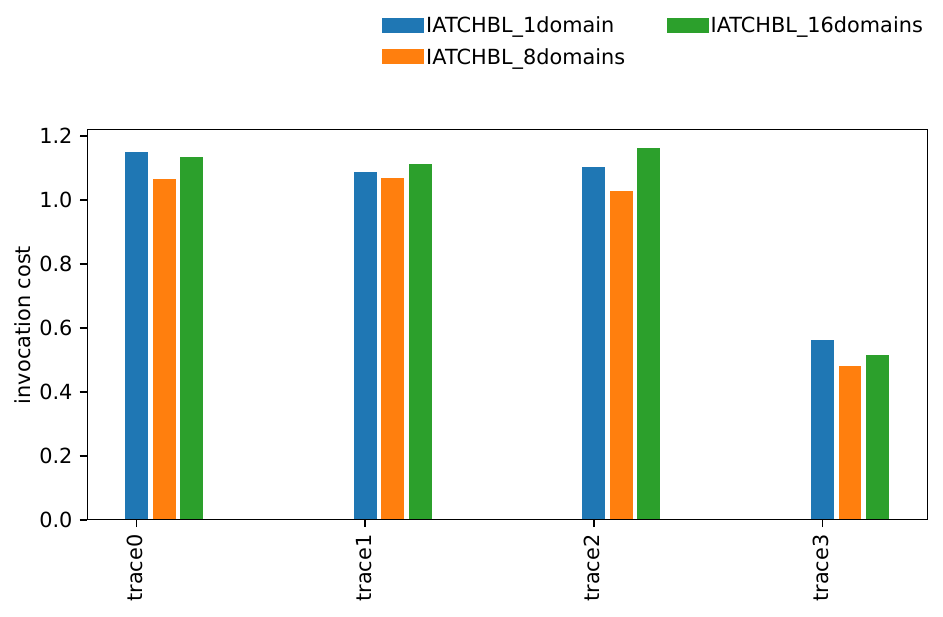}
    \caption{Workload: invocation cost}
    \label{fig:eval:summary:cost}
\end{figure}

The latency and invocation cost of the requests has increased by $\approx10\%$ for under and medium utilized system - Figure-\ref{fig:eval:summary:E2EMean}, Figure-\ref{fig:eval:summary:cost}. While it has reduced by $50\%$ for over-utilized system. An over-utilized system has amply filled run queues (722 1-min loadaverage) that gives custom scheduler enough space for optimization. Overall, 8domain configuration performs best with only $5\%$ increase in latency and invocation cost. 

\begin{figure}[htbp] % Use figure* for full page width (two-column documents)
    \centering
    \includegraphics[width=0.4\textwidth]{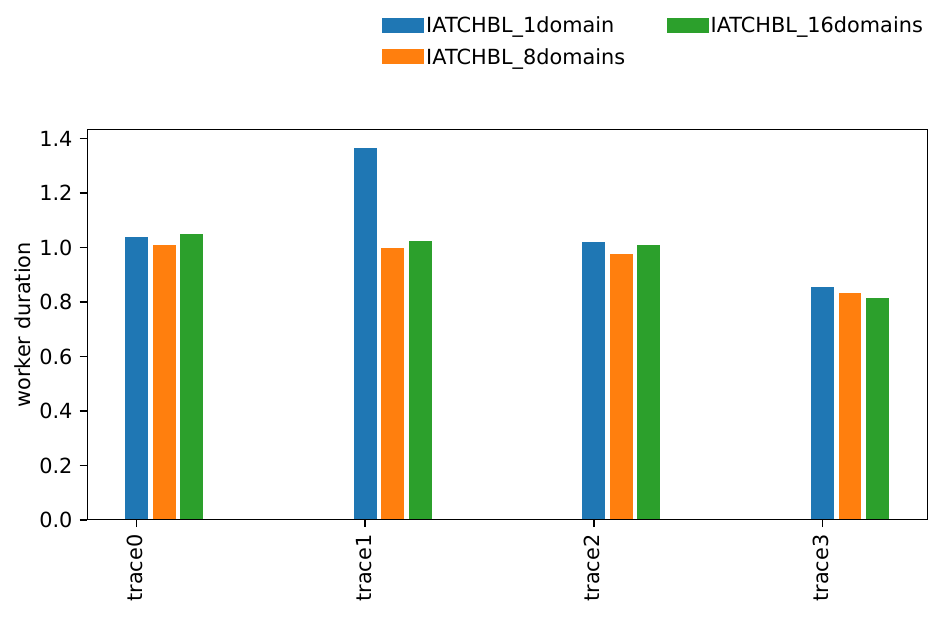}
    \caption{Workload: worker duration (excludes queue time)}
    \label{fig:eval:summary:WorkerDur}
\end{figure}

The workerduration which excludes the queue time increases by $5\%$ for under and mediumly utilized system. While it reduces by $15\%$ for over utilized system - Figure-\ref{fig:eval:summary:WorkerDur}. The single queue loadbalancing increases the probability of a task finding a high frequency core. This effect is clearly visible in over utilized system because there are enough tasks in queue to take advantage of a warm core. For 8domain configuration there is virtually no increase in worker duration. Instead, it reduces by $5\%$ for trace2 which is a medium utilized system.

Overall dividing the system into eight domains give best results. It reduces energy usage by $\approx 15\%$ while increasing invocation cost by $\approx 5\%$. It strikes the balance between keeping functions separate to avoid neighbor effect and taking advantage of single queue loadbalancing cum prioritization within a scheduling domain.

%% file: related_work.tex
\section{Related Work}

Driving CPU scheduling from the serverless control plane has been explored by Tailclipper, SFS and ALPS \cite{tailclipper}\cite{sfs}\cite{alps}. Tailclipper prioritizes tasks based on arrival time of the request to avoid reordering of tasks and stagglers. SFS and ALPS prioritizes tasks based on execution time of functions to implement shortest remaining processing time scheduler. None of these explore loadbalancing the tasks among scheduling domain based on function characteristics. 

Shenango, Shinjuku and Caladan are targeted solutions to optimize CPU scheduling for data plane applications \cite{shenango}\cite{shinjuku}\cite{caladan}. Nest is a generic scheduler that aims to keep tasks together on few cores to improve the performance and efficiency of the system as a whole\cite{nest}. These solution either solely focus on scheduler or it's application in data plane context instead of serverless.

Autopilot, Sinan, Kubernetes autoscalar and FIRM adjust the control group cpu bandwidth limits to pack more functions on a server \cite{autopilot}\cite{sinan}\cite{k8sautoscale}\cite{firm}. These solution do not optimize the CPU scheduling of tasks of the functions.

%% file: conclusion.tex
\section{Conclusion}

This paper demonstrates that integrating CPU scheduling with the serverless control plane enables new optimization opportunities beyond those available in traditional operating system schedulers. By exposing a lightweight interface between the control plane and a custom eBPF-based SchedExt scheduler, our approach allows scheduling domains to be dynamically assigned using workload characteristics such as function inter-arrival time. The proposed scheduler combines domain-aware load balancing with single-queue scheduling and virtual-time prioritization to improve core reuse and reduce interference from busy-polling tasks.

Our implementation in the Ilúvatar serverless platform shows that the approach is practical and incurs minimal scheduling overhead. Experimental evaluation using Azure workload traces \cite{azuretrace} demonstrates that an eight-domain configuration provides the best overall balance, reducing energy consumption by approximately 15\% while increasing invocation cost by only 5\%. Furthermore, under highly utilized workloads, the scheduler significantly improves performance by reducing latency by up to 50\% compared to Linux CFS. Although workloads containing long-running CPU-bound applications may experience modest overheads, such workloads are uncommon in serverless environments.

Overall, these results show that serverless platform-driven CPU load balancing is a promising direction for improving the efficiency and performance of modern FaaS systems. Future work includes developing adaptive domain sizing strategies, incorporating additional workload characteristics into scheduling decisions, and extending the framework to jointly optimize CPU, memory, and heterogeneous accelerator resources.